\documentclass[prd,aps,amssymb,amsmath]{revtex4}
\usepackage{graphicx}% Include figure files
\usepackage{bm}% bold math
\begin{document}

\title{SuperGZK neutrinos: testing physics beyond the Standard Model}

\author{Veniamin Berezinsky   }
 \affiliation{INFN, Laboratori Nazionali del Gran Sasso, I-67010 
  Assergi (AQ), Italy\\ 
and  Institute for Nuclear Research of the RAS, Moscow, Russia}

\begin{abstract} 
The sources and fluxes of superGZK neutrinos, $E>10^{20}$~eV, 
are discussed. The most promising sources are reionization bright phase, 
topological defects, 
superheavy dark matter and mirror matter. The energy of neutrinos can
be above the GUT scale ($\sim 10^{16}$~GeV). The predicted fluxes are 
observable by future space detectors EUSO and OWL.
\end{abstract}
\maketitle
\section{Introduction}

The abbreviation `SuperGZK neutrinos' implies neutrinos with energies
above the Greisen-Zatsepin-Kuzmin \cite{GZK} cutoff 
$E_{\rm GZK} \sim 5\times 10^{19}$~eV. Soon after theoretical
discovery of the GZK cutoff it was noticed that this phenomenon is 
accompanied by a flux of UHE neutrinos, that in some models can be
very large \cite{BZ69}. In 80s it was realized that 
topological defects can
produce unstable superheavy particles with
masses up to the  GUT scale \cite{Witten} and neutrinos with
tremendous energies  can emerge in this process \cite{HiSch}. 

It has been proposed that SuperGZK 
neutrinos can be detected with help of horizontal Extensive Air Showers  
(EAS) \cite{BeSm}. 
The exciting prospects for detection of SuperGZK neutrinos have appeared 
with the ideas of space detection, e.g. in the projects 
EUSO\cite{EUSO} and OWL\cite{OWL}. The basic idea of detection can
be explained by example of EUSO. 

A superGZK neutrino entering the Earth atmosphere in near-horizontal 
direction produces an EAS. The known fraction of its energy, which
reaches 90\% , is radiated in form of isotropic fluorescent light. 
An optical telescope from a space observatory detects this
light. Since the observatory is located at very large height 
($\sim 400$~km) in comparison with thickness of the atmosphere, the
fraction of detected flux is known, and thus this is the calorimetric 
experiment (absorption of light in the upward direction is 
small). A telescope with diameter 2.5 m controls the area 
$\sim 10^5$~km$^2$ and has a threshold for EAS detection 
$E_{\rm th}\sim 1\times 10^{20}$~eV.     

\section{Sources of SuperGZK neutrinos}
The sources can be of the accelerator and non-accelerator origin.
\\*[2mm]
{\it Accelerator sources}\\*[2mm]
\noindent
A source comprises an {\it accelerator} and {\it target}. 
Low energy photons inside the source or outside it (e.g. CMB photons)
are usually more efficient for neutrino production than gas. The
problem is acceleration to 
$E \gg 1.10^{20}$~eV. For shock acceleration the maximum energy can
reach optimistically $E_{\rm max} \sim 10^{21}$~eV for protons. There
are some other, less developed accelerator mechanisms (e.g. 
unipolar induction, acceleration in strong e-m waves), which hopefully
can provide larger $E_{\rm max}$. An interesting mechanism with
very large $E_{\rm max}$ has been  recently proposed in 
Ref.\cite{Chen}.    

The large neutrino fluxes at the superGZK energies is also a problem 
for accelerator sources.\\*[2mm]  
\noindent
{\it Non-accelerator (top-down) sources}\\*[2mm]
\noindent
{\it Topological Defects} (TD) and {\it Superheavy Dark Matter} (SHDM) can
easily produce
required superGZK energies. In both cases the superGZK neutrinos are
produced in the decays of very heavy particles, unstable gauge and 
Higgs particles in case of TDs, and quasistable particles in case of SHDM.   
Neutrinos are born mostly in pion decays and have 
$E_{\rm max} \sim 0.1 m_X$. A natural upper limit for mass  is 
$m_X \leq m_{\rm GUT} \sim 10^{16}$~GeV.   

TDs differ substantially by mechanisms of SH particle production. 

They include emission of X-particles through cusps in {\it superconducting 
cosmic strings}. In this case the energy of X-particle is boosted by 
the Lorentz-factor of the cusp which can reach $\Gamma \sim 10^3 -
10^6$. 

In case of {\it network of monopoles} connected by strings, monopoles
move with large acceleration $a$ and can thus radiate the  
gauge bosons such as gluons, $W^{\pm},Z^0$. Neutrinos appear in their
decays, in case of gluons through production and decays of pions. 
The typical energy of radiated quanta is $E \sim a \Gamma_M$, 
where $\Gamma_M$ is the Lorentz-factor of the monopole. 

In {\it necklaces}, where each monopole (antimonopole) is attached 
to two strings, 
all monololes and antimonopoles
inevitably annihilate in the end of evolution, producing neutrinos
in the this process (see Subsection \ref{subsec:neckl}). 

{\it Vortons} can decay due to quantum tunnelling, producing neutrinos. 

In some of the cases listed above the neutrino energies can be larger
(due to Lorentz factor) than the GUT scale.

{\it SHDM} is very efficiently produced at inflation due to
gravitational radiation and is accumulated in galactic halos with  
overdensity $\sim 10^5$. Neutrinos are produced at the decay of 
these particles 
mostly in extragalactic space. 
(see Subsection \ref{subsec:SHDM}).

For more details of the top-down particle production see reviews \cite{td}.
\section{Upper limits on diffuse neutrino fluxes}
%There are different upper bounds for the diffuse neutrino flux.  
{\it Cascade upper limit \cite{BeSm,book}}\\*[2mm]
\noindent
Production of HE neutrinos is accompanied by photons and electrons
which start e-m cascade colliding with target photons
($\gamma+\gamma_{\rm tar} \rightarrow e^++e^-$,~ ~
$e+\gamma_{\rm tar}\rightarrow e'+\gamma'$). The cascade photons get 
into range observed by EGRET, and the energy density of these photons,
$\omega_{\rm cas}$, put the upper limit on diffuse  HE neutrino flux 
$J_{\nu}(E)$ as it follows from the obvious chain of inequalities 
\cite{BeSm,book}:
$$
\omega_{\rm cas}>\omega_{\nu}(E)\equiv\frac{4\pi}{c}\int_E^{\infty}
EJ_{\nu}(E)dE>\frac{4\pi}{c}E\int_E^{\infty}J_{\nu}(E)dE=
\frac{4\pi}{c}EJ_{\nu}(>E),
$$
from which the bound for differential neutrino flux follows
\begin{equation}
E^2I_{\nu}(E) \leq (c/4\pi)\omega_{\rm cas}.
\label{eq:cas}
\end{equation}
The upper bound Eq.~(\ref{eq:cas}) is obtained under assumptions that
neutrinos are produced in the chain of decays of particles 
(e.g. pions, $W$ and $Z^0$), where electrons or photons appear too, 
and that the sources are not opaque for the cascade photons. In fact cascade
can develop inside a source, provided that produced cascade photons 
are not absorbed by gas in the source. Only fully opaque (`hidden`)
sources escape the bound Eq.~(\ref{eq:cas}), e.g. see 
Subsection \ref{subsec:mirror}. 

This bound has a  great 
generality. It is valid for photon and proton target and for any
spectrum index (in this case the bound becomes up to factor 2 stronger
\cite{book}, p.359). \\*[2mm]
{\it Waxman-Bahcall (WB) upper limit \cite{WB}}\\*[2mm]
\noindent
If UHE protons escape freely from a source, their flux should
be less than the observed one. It gives the WB bound on neutrino flux produced
by escaping protons. It is obtained for a specific shape ($1/E^2$) of
the production spectrum. This bound is much stronger than the cascade
limit, but it is not valid for many sources and models (e.g. for TDs and
 SHDM, where production of protons is negligible, for acceleration
sources at high redshift, from which UHE protons do not reach us, for 
clusters of galaxies, since time of CR exit from there is larger than
age of universe, for some specific AGN models, e.g. the Stecker model etc).
However, it is valid for diffuse flux from some interesting 
sources like GRBs and some models of AGN.\\*[2mm] 
{\it Mannheim-Protheroe-Rachen (MPR) upper bound \cite{MPR}}\\*[2mm]
\noindent
As compared to the WB bound, the MPR limit is valid for sources with 
various optical depths and for different maximal acceleration
energies. It is weaker than WB bound, and for high and low energies
differ not too much from the cascade limit.\\*[2mm]
\noindent
Since UHE protons is subdominant component for most sources of
superGZK neutrinos (see next Section), the WB and MPR bounds are not valid 
for them, and the cascade bound becomes most appropriate.

\section{Diffuse fluxes of superGZK neutrinos}\label{sec:fluxes}

\subsection{AGN and other accelerator sources}\label{subsec:AGN}
The fluxes of UHE neutrinos produced in $p\gamma$ collisions with 
CMB photons by UHE protons from AGN (quasars and Seyfert galaxies) 
were calculated in Ref.\cite{BeSm} up to $E_{\rm max} \sim 10^{21}$ eV.
The assumed cosmological evolution of the sources up to redshift 
$z=14$ with the evolutionary index $m=4$ increases strongly the flux.
The cascade upper bound has been taken into account. 

Recently the detailed  calculations of diffuse neutrino fluxes from 
evolutionary and non-evolutionary sources have been performed 
in Ref.\cite{Kal} (see also \cite{Fod}). The acceleration mechanisms
are not specified, but
it is assumed that generation spectra are flat and $E_{\rm max}$
can reach $3\times 10^{22}$~eV. Neutrinos are produced in collisions
with CMB photons. 
The calculated fluxes agree by order
of magnitude with those from Ref.\cite{BeSm} 
\begin{figure}[htb]
\begin{center}
\includegraphics[width=10cm,height=9cm]{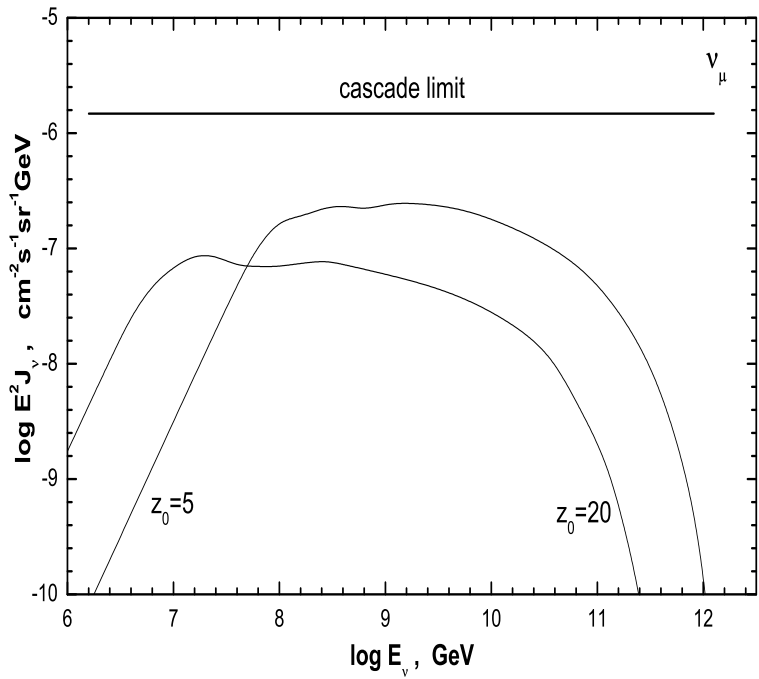}
\end{center}
%\centerline{\epsfxsize=3.7in\epsfbox{procs-bright.eps}}   
\caption{Diffuse $\nu_{\mu}$ neutrino flux from the reionization
bright phase at redshift $z_0=20$ and $z_0=5$ \protect\cite{BeGa}. 
The cascade upper bound for
$\nu_{\mu}$ neutrinos is shown by curve `cascade limit'.
\label{bright}}
\end{figure}
\subsection{Reionization bright phase}\label{subsec:bright}
The WMAP detection of reionization \cite{WMAP} implies an early
formation of stars at large
redshifts up to $z \sim 20$, able to reionize the universe. 

A plausible process is formation of very
massive, $M> 100 M_{\odot}$, Population III stars with subsequent 
fragmentation to SN. 
Two-burst scenario of reionization is plausible:  at      
$z \approx 15$ and $z \approx 6$ \cite{Cen}. The fraction of baryonic
matter in the form of compact objects is $\sim 0.01$ at 
$z \sim 10$ \cite{WMAP}. 
Fragmentation of massive Pop III stars into presupernovae  and 
black holes results in CR acceleration by various mechanisms
(shocks, jets in miniquasars and hypernovae etc). Not 
specifying them we shall assume that the energy release in the form of CR
is $W_{\rm cr} \sim 5\times 10^{50}$~erg/$1M_{\odot}$, generation spectrum 
is $\sim E^{-2.1}$ and $E_{\rm max}= 1\times 10^{13}$~GeV. 
Neutrinos are produced in collisions with CMB photons.
In Fig.\ref{bright} we present the calculated neutrino spectrum for two
values of $z$~ (20 and 5) \cite{BeGa},
together with cascade upper limit for one neutrino
flavour. This model is very similar to the galactic bright phase 
(for a review see \cite{book}, p.352). 
\begin{figure}[htb]
%\epsfxsize=10cm   %width of figure - will enlarge/reduce the figures
%\epsfbox{fig3.eps}
%\figurebox{2cm}{3cm}{} %to have a box alone 
\begin{center}
\includegraphics[width=11cm,height=7cm]{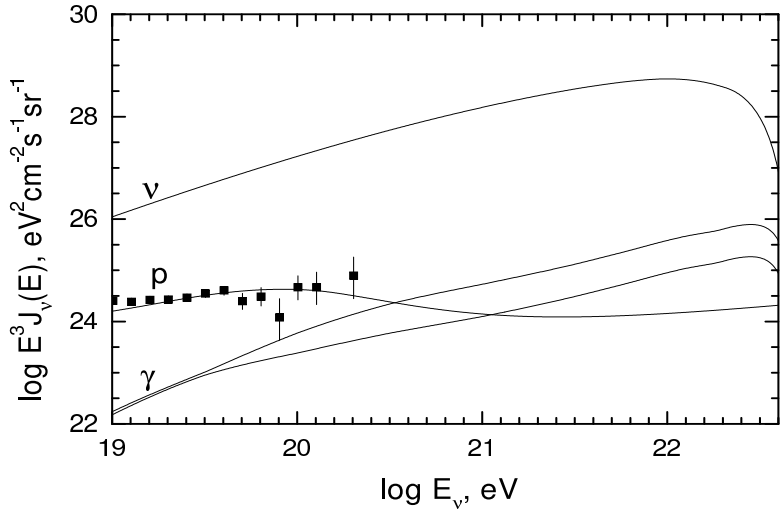}
\end{center}
%\centerline{\epsfxsize=3.7in\epsfbox{procs-neckl.eps}}   
\caption{Diffuse neutrino flux (curve $\nu$) from 
necklaces \protect\cite{ABB}. The curve 
$p$ shows the UHE proton flux, compared with the AGASA data,
and two curves $\gamma$ show UHE photon flux for two cases of absorption.
\label{necl}}
\end{figure}
\vspace{-5mm}
\subsection{Necklaces}\label{subsec:neckl}
Necklaces are hybrid TDs (monopoles connected by strings) formed in a
sequence of symmetry breaking phase transition 
$G\rightarrow H\times U(1)\rightarrow H \times Z_2$. In the first
phase transition the monopoles are produced, and at the second each
monopole gets attached to two strings. 

The symmetry breaking scales of these two phase transitions, $\eta_m$
and $\eta_s$, are the main parameters of the necklaces. 
They give the monopole mass, $m\sim 4\pi \eta_m/e$, and tension of the string,
$\mu \sim 2\pi \eta_s^2$. 
The evolution of necklaces is governed by
ratio $r=m/\mu d$, where $d$ is a length of a string between two
monopoles. During the evolution this length diminishes due to
gravitational radiation, and in the end all monopoles and
antimonopoles annihilate, producing high energy neutrinos as the
dominant radiation. The diffuse neutrino flux from necklaces is given
\cite{ABB} in Fig. \ref{necl}.
Note, that the rate of UHE particle production in this model is 
$\dot{n}_X \sim r^2\mu/t^3m_X$, where $r$ tends 
in evolution of a necklace to 
$r_{\rm max} \sim \eta_m/\eta_s \sim 10^{13}$ contrary to expectation in 
Ref.\cite{Kal} that this parameter is typically $\sim 1$ for all TDs.
%\vspace{-4mm}
\begin{figure}[htb]
%\epsfxsize=10cm   %width of figure - will enlarge/reduce the figures
%\epsfbox{fig3.eps}
%\figurebox{2cm}{3cm}{} %to have a box alone 
\begin{center}
\includegraphics[width=13cm,height=8cm]{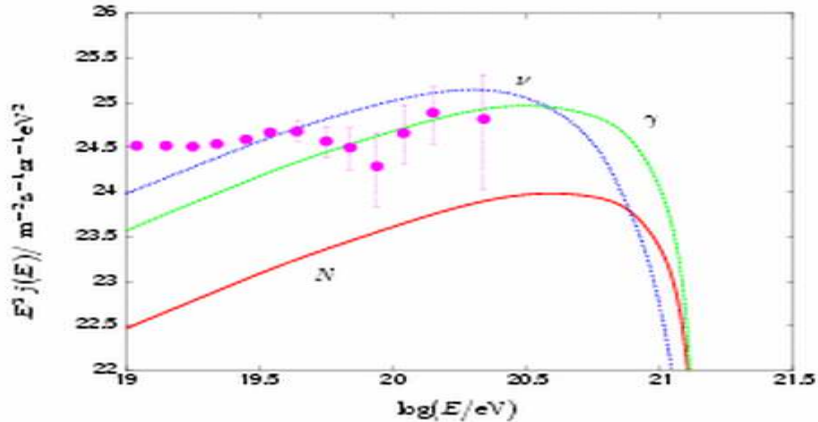}
\end{center}
%\centerline{\epsfxsize=3.7in\epsfbox{procs-neckl.eps}}   
\caption{Diffuse neutrino flux (curve $\nu$) from 
SHDM \protect\cite{BKV,BK-MC}.
The curve $\gamma$ shows UHE photon flux from the halo compared with 
the AGASA data, and 
the curve 
$N$ shows the UHE nucleon flux from the halo.
\label{SHDM}}
\end{figure}
%\vspace{-10mm}
\subsection{Superheavy dark matter (SHDM)}\label{subsec:SHDM}
Production of SHDM particles naturally occurs in time varying
gravitational field of the expanding universe at post-inflationary
stage \cite{KuTk,Kolb}. This mechanism of production does not depend 
on the coupling of X-particles with other fields, and occurs even in 
case of sterile X-particles. 
The relic density of these particles is mainly determined 
(at fixed reheating temperature and inflaton mass) by
mass $m_X$. SHDM can constitute all CDM observed in the universe,
or be its subdominant component.
The range of practical interest is given by masses $(3 - 10)10^{13}$~GeV, 
at larger masses the SHDM is subdominant. 

SHDM is accumulated in the Galactic halo with overdensity $\sim 10^5$. 

In many elementary-particle models SH particles can be quasi-stable
with lifetime $\tau_X \gg 10^{10}$~yr. Such decaying particles produce UHECR 
with photons from the halo being the dominant component. The measured
flux of these photons with corresponding signatures (anisotropy in the
direction of Galactic Center and the muon-poor EAS) determines $\tau_X$
experimentally. {\it Such precise  determination of a parameter from 
experimental data has nothing to do with fine-tuning}.

The energy spectrum of HE particles is now reliably calculated in 
the SUSY-QCD framework as $\sim dE/E^2$, there is an agreement between
the calculations of different groups \cite{BK-MC,SaTo,BaDr}.

Neutrino flux from SHDM is given in Fig.\ref{SHDM} according to 
Refs.\cite{BKV,BK-MC}. It is
produced by decays of X-particles in extragalactic space, while UHECR
signal is produced mainly by photons from X-particle decays in the
Galactic halo.
\subsection{Mirror matter}\label{subsec:mirror}
Mirror matter can be most powerful source of UHE neutrinos not limited
by the usual cascade limit \cite{mirror}. 

Existence of mirror matter is an interesting theoretical idea which
was  introduced by Lee and Yang (1956), Landau (1957) and most notably
by Kobzarev, Okun and Pomeranchuk (1966) to have a space reflection
operator $I_s$ commuting with Hamiltonian $[I_s,H]=0$. The mirror 
particle space is generated by operator $I_s$, which transfers 
the left states of ordinary particles into right states of the mirror 
particles and vise versa. The mirror particles have interactions identical 
to the ordinary particles, but these two sectors interact with each other
only gravitationally. Gravitational interaction mixes the visible 
and mirror neutrino states, and thus causes the oscillation between
them.  

A cosmological scenario must provide the suppression of the mirror
matter and in particular the density of mirror neutrinos at the epoch 
of nucleosynthesis. It can be obtained in the two-inflaton model 
\cite{mirror}. The rolling of two inlaton to minimum of the potential
is not synchronised, and when mirror inflaton reaches minimum, the ordinary
inflaton continues its rolling, inflating thus the mirror matter produced
by the mirror inflaton. While mirror matter is suppressed, the mirror
topological defects can strongly dominate \cite{mirror}. Mirror TDs 
copiously produce mirror neutrinos with extremely high
energies typical for TDs, and they are not accompanied by
any visible particles. Therefore, the upper limits on HE mirror 
neutrinos in our world do not exist. All HE mirror particles 
produced by mirror TDs are sterile for us, interacting with
ordinary matter only gravitationally, and only mirror neutrinos 
can be efficiently converted into ordinary ones due to oscillations. 
The only (weak) upper limit comes from the resonant
interaction  of converted neutrinos with DM neutrinos: 
$\nu+\bar{\nu}_{\rm DM}\to Z^0$ \cite{mirror}.

\begin{figure}[ht]
%\epsfxsize=10cm   %width of figure - will enlarge/reduce the figures
%\epsfbox{fig3.eps}
%\figurebox{2cm}{3cm}{} %to have a box alone 
\begin{center}
\includegraphics[width=11cm,height=8cm]{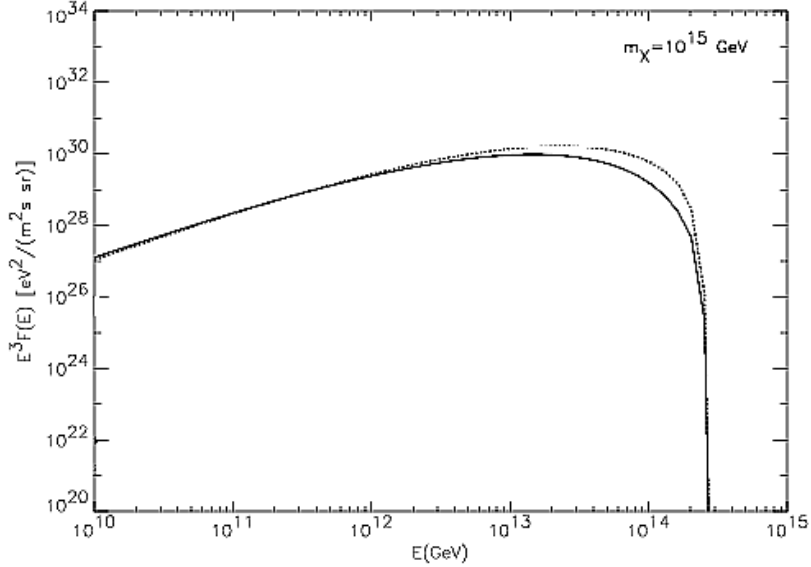}
\end{center}
%\centerline{\epsfxsize=3.5in\epsfbox{procs-mirror.eps}}   
\caption{Diffuse neutrino flux from mirror necklaces \protect\cite{ABB}. 
\label{mirror}}
\end{figure}

In Fig.\ref{mirror} UHE neutrino flux from mirror TDs is presented 
according to calculations \cite{ABB}. This model provides the largest
flux of superGZK neutrinos being not restricted by the standard cascade
bound (\ref{eq:cas}).
\section{Conclusions}
The detectable fluxes of superGZK neutrinos ($E_{\nu} > 10^{20}$~eV)
can be produced by accelerator and non-accelerator sources. 

The accelerator sources need $E_{\rm max} \gg 1\times 10^{21}$ eV,
which most probably implies the non-shock acceleration mechanisms. 

The non-acceleration sources, topological defects and superheavy relic 
particles, involve physics beyond the Standard Model. Very high
neutrino energies, in excess of the GUT scale, are possible for these
sources. The fluxes are normally limited by the cascade bound. The 
largest flux (above the standard cascade bound) is predicted by the mirror
matter model.

\end{document}